\documentclass[preprint,epsfig,onecolumn,floats, showkeys]{revtex4}
\usepackage{makeidx}
\usepackage{amsmath}
\usepackage{amssymb}
\usepackage{graphicx}
\usepackage{epsfig}

\begin{document}

\title{Quantum Pattern Recognition of Classical Signal}
\author{Chao-Yang Pang}
\email{cyp_900@hotmail.com} \email{cypang@sicnu.edu.cn}
\affiliation{Sichuan Key Lab of Software, Sichuan Normal University,
Chengdu 610068, P.R. of China} \affiliation{College of Mathematics
and Software Science, Sichuan Normal University, Chengdu 610068,
P.R. of China}
\author{Cong-Bao Ding}
\affiliation{College of Physics and Electronic Engineering, Sichuan Normal University,
Chengdu 610068, P.R. of China}
\author{Ben-Qiong Hu}
\affiliation{College of Information Management, Chengdu University of Technology, 610059,
P.R. of China}

\begin{abstract}
It's the key research topic of signal processing that recognizing genuine
targets real time from the disturbed signal which has giant amount of data.
A quantum algorithm for pattern recognition of classical signal which has
time complexity $O(\sqrt{N})$ is presented in this paper.
\end{abstract}

\keywords{ Pattern recognition, Grover's algorithm, Rotation on subspace}
\maketitle

\section{.Introduction}

It's the key research topic of signal processing that recognizing
genuine targets real time from the disturbed signal which has giant
amount of data. A quantum pattern recognition of classical signal is
presented. To understand the idea of this paper easily, the example
of designing quantum pattern recognition to detect the saturated
raid from armada is presented, and all contents are focused on this
example.

Saturated raid is the topical raid mean of an armada, which many missiles,
many planes, many spurious weapons generated by electromagnetic wave, and
et. al. will be appeared at a same. Defender's first task is to recognize
the genuine space targets \emph{real time} from the signal which is captured
by phased array radar. This is a hard problem for classical computer. Peter
Shor's quantum algorithm \cite{Shor} and Grover's quantum search algorithm
\cite{Grover} show the high efficiency of quantum computer. Is there fast
quantum algorithm to detect the saturated raid from an armada and recognize
genuine space targets at real time?

Pang presented a quantum loading scheme (QLS) \cite{QLS} to make quantum
computer compatible with classical memory. The QLS can load all giant signal
data captured by phased array radar into quantum registers at a time.
Quantum image match algorithms and the method rotation on subspace show that
image recognition is possible \cite{QIC, QVQ1, QDCT, QVQ2}. Quantum discrete
Fourier transform with classical output (QDFT) shows that quantum digital
signal processing (especially for radar signal) is possible \cite{QDFT}.

\textbf{Introduction of Quantum Loading Scheme }$U_{L}$ \cite{QLS}\textbf{: }%
Suppose an arbitrary record (or data) $record_{i}$ is stored in classical
database with the corresponding index $i$. The function of unitary$\ $%
operation $U_{L}$ can be described as
\begin{equation}
\frac{1}{\sqrt{N}}({\sum\limits_{i=0}^{N-1}{{{\left\vert {i}\right\rangle
\left\vert 0\right\rangle )\left\vert {ancilla}\right\rangle }}}}\overset{%
U_{L}}{{\rightarrow }}\frac{1}{\sqrt{N}}({\sum\limits_{i=0}^{N-1}{{{%
\left\vert {i}\right\rangle \left\vert record_{i}\right\rangle )\left\vert {%
ancilla}\right\rangle }}}}  \label{eqUL}
\end{equation}

, where ancillary state ${{{{\left\vert {ancilla}\right\rangle }}}}$ is
known.

That is, unitary$\ $operation $U_{L}$ loads all information of records
stored in a classical database into quantum state. Unitary$\ $operation $%
U_{L}$ has time complexity $O(logN)$ (unit time: phase transform and
flipping the qubits of registers) \cite{QLS}. Operator $U_{L}$ is so fast
that its running time can be ignored when analyzing the time complexity of a
algorithm.

\textbf{Introduction of Quantum Search Algorithm with Complex Computation
(i.e., the Method of Rotation at Subspace) \cite{QIC, QVQ1, QDCT, QVQ2}:}

Grover's algorithm can find a database record according to the given index.
However, database search is complex in general. E.g., police often hopes to
find a mug shot from the database in which many sample photos are stored by
the method of matching every sample photo and the photo captured by the
vidicon at the entrance of airport at real time. Grover's algorithm is
invalid for this kind of search case because the coupling between search and
other computation (e.g., image matching) is required at this case. Pang
et.al. presents a quantum method named "\textbf{rotation at subspace}" \cite%
{QIC, QVQ1, QDCT, QVQ2} to generalize Grover's algorithm to the search case
with arbitrary complex computation, that is derived from the research of
quantum image compression \cite{QIC}. The method of rotation at subspace is
described as following briefly:

First, All input datum are stored in classical memory as database records.
Assume that total number of records is $N$. All these records can be loaded
into a superposition of state using quantum loading scheme $U_{L}$.

Second, construct the \textbf{general Grover iteration (GGI)} $G_{general}$
as

\begin{equation}
G_{general}=(2|\xi \rangle \langle \xi |-I)(U_{L})^{\dag }(O_{c})^{\dag
}O_{f}O_{c}U_{L}  \label{eqGeneralGroverIteration}
\end{equation}

, where $O_{c}$ denotes computation oracle such as image matching, $f\ $%
denotes the judge function (i.e., if the output of $O_{c}$ satisfies some
conditions, let $f=1$, else $f=0$),$\ O_{f}$ is the oracle of the judge
function, and $|\xi \rangle =\frac{1}{\sqrt{N}}\sum\limits_{i=0}^{N-1}|i%
\rangle $.

Third, similar to Grover's algorithm, let unitary operation $G_{general}$
act on initial state $\frac{1}{\sqrt{N}}(\underset{i=0}{\overset{N-1}{\sum }}%
{\left\vert {i}\right\rangle }_{{registers1}}){\left\vert 0\right\rangle }%
_{registers2}$ $O(\sqrt{N})$ times, we will find the optimal solution.

\section{The Proposed Architecture of Multi-Pattern Recognition to Detect
Saturated Raid}

Target pattern set is denoted by
\begin{equation*}
P=\{t_{i}|i=0,1,...,n-1\}
\end{equation*}%
, where $n$ is the number of elements in $P$ and $t_{i}$ is the information
of target pattern, such as image captured by radar.

Spurious pattern set is denoted by%
\begin{equation*}
S=\{s_{j}|j=0,1,...,s-1\}
\end{equation*}%
, where $s$ is the total number of elements in $S$ and $s_{j}$ is the
information of spurious pattern which is generated by enemy's
electromagnetic wave.

We often have $s+n\neq 2^{x}$, where $x$ denotes some integer. If $s+n\neq
2^{x}$, add some virtual patterns $v\in V$ for the condition $|P\cup S\cup
V|=2^{x}$ to simplify computation.

Let the whole set
\begin{equation*}
I=P\cup S\cup V
\end{equation*}
, and the number of elements of set $I$ is $N$.

We always depend on feature to recognition target pattern in set $P$. Assume
that all characteristics are collected as set%
\begin{equation*}
C=\{c_{-2},c_{-1},c_{0},c_{1},\cdots ,c_{m-1}\}
\end{equation*}%
, where $m\leq n$ in general.

The computation of the element of set $C$ depends on the computation method
of feature. E.g., we can calculate out the\ affine and projective invariants
\cite{HGeometry} (or other topology invariants) of bomb carrier or aircraft
carrier as the pattern characteristics, which are \emph{not} changed by
disturbing signal in theory.

The mapping between pattern set $I$\ and feature $C$ (i.e., $g:I\mapsto C$)
is denoted as%
\begin{equation*}
g(x_{i})=\left\{
\begin{tabular}{ccc}
$c_{i_{k}}$ & $,$ & $x_{i}\in P$ \\
$c_{-1}$ & $,$ & $x_{i}\in S$ \\
$c_{-2}$ & $,$ & $x_{i}\in V$%
\end{tabular}%
\right.
\end{equation*}%
, where $0\leq i_{k}<m$. The feature of spurious pattern (i.e., $x_{i}\in S$%
) is denoted by $c_{-1}$ and the feature of the meaningless virtual pattern
(i.e., $x_{i}\in V$)\ is denoted by $c_{-2}$.

We often have a sample set (or codebook) about all target patterns
previously, which is stored as a database \cite{QIC, QVQ1, QDCT, QVQ2, PR}.
Assume that the sample set is
\begin{equation*}
CB=\{(t_{i},c_{i_{k}})|k=0,1,...,m-1\}
\end{equation*}%
, where $c_{i_{k}}$ is the feature of target pattern.

The whole quantum multi-pattern recognition proposed in this paper is
described as follows:

First, radar captures the signals and they are stored in memory, denoted as
set $P\cup S$ . Store all datum of set $I=P\cup S\cup V$ in classical memory
with indices as a database.

Second, load all information of data in set $I$ into a state using QLS.

Third, for a given feature stored in $CB$, design the following quantum
algorithm to judge if the feature is hidden in the state and recognize it.
Using the same method to full search the sample set $CB$, all target
patterns will be recognized.

\begin{figure}[tbh]
\epsfig{file=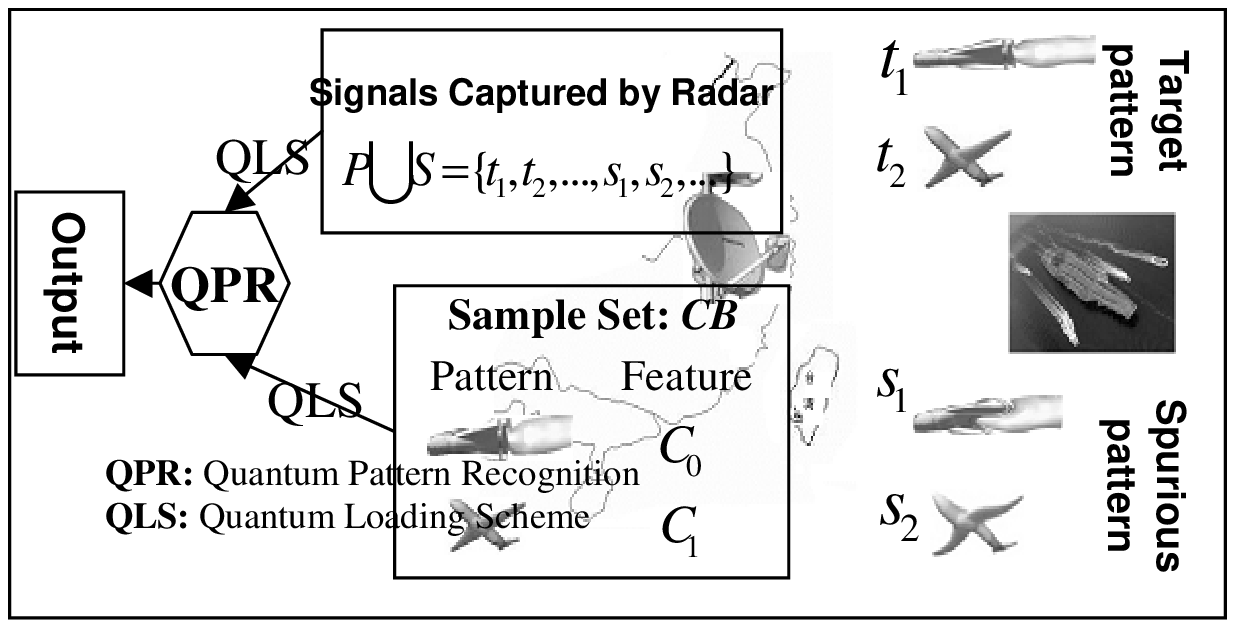,width=11cm,} \caption{\textbf{The Architecture
of Quantum Multi-Pattern Recognition}} \label{figPR}
\end{figure}

Fig\ref{figPR} illustrates the architecture of multi-pattern recognition
proposed by this paper.

\section{Quantum Multi-Pattern Recognition}

\textbf{First, construct the following data structure (DS) and unitary
operations :}

\textbf{DS1.} Save all elements of set $I$ in classical memory, and these
records have indices $i=0,1,...,N-1$.

\textbf{DS2.} Construct six registers to denote the information of the
element in set $I$. The six registers have data format
\begin{equation*}
|\alpha \rangle _{register1}|c_{i_{0}}\rangle _{register2}\otimes |i\rangle
_{register3}\otimes |x_{i}\rangle _{register4}\otimes |c_{i}\rangle
_{register5}\otimes |d(c_{i},c_{i_{0}})\rangle _{register6}
\end{equation*}

\ \ \ \ That is, 1st, 2nd, 3rd, 4th, 5th, and 6th register are used to save
the input parameter $\alpha $, the pattern characteristics $c_{i_{0}}\in CB$
which required to be extracted from the superposition of state, the index of
the element of $I$, pattern $x_{i}\in I$, the feature of pattern $x_{i}$,
and the similarity measurement value between $c_{i_{0}}$ and $c_{i}$.

\ \ \ \ Let initial state is
\begin{equation*}
|\psi _{0}\rangle =|\alpha \rangle |c_{i_{0}}\rangle |0\rangle |0\rangle
|0\rangle |0\rangle
\end{equation*}

\ \ \ \ Hardmard transform $H$ acts on $|\psi _{0}\rangle $ will generate
the following state:
\begin{equation*}
|\psi _{0}\rangle \overset{H}{\rightarrow }|\psi _{1}\rangle =\frac{1}{\sqrt{%
N}}|\alpha \rangle |c_{i_{0}}\rangle (\overset{N-1}{\underset{i=0}{\sum
|i\rangle }})|0\rangle |0\rangle |0\rangle
\end{equation*}

\textbf{DS3.} Construct loading operation $U_{L}$ as \cite{QLS}
\begin{equation*}
|\alpha \rangle |c_{i_{0}}\rangle |i\rangle |0\rangle |0\rangle \overset{%
U_{L}}{\rightarrow }|\alpha \rangle |c_{i_{0}}\rangle |i\rangle
|x_{i}\rangle |0\rangle |0\rangle
\end{equation*}

\ \ \ \ All information of set $I$ will be loaded into registers by
operation $U_{L}$.

\textbf{DS4.} Design oracle $O_{c}$ to compute the pattern feature, i.e.,

\begin{equation*}
|\alpha \rangle |c_{i_{0}}\rangle |i\rangle |x_{i}\rangle |0\rangle \overset{%
O_{c}}{\rightarrow }|\alpha \rangle |c_{i_{0}}\rangle |i\rangle
|x_{i}\rangle |c_{i_{k}}\rangle |0\rangle
\end{equation*}

\ \ \ \ If the computation of oracle $O_{c}$ is complex, decompose it as
many simple oracles. That is supported by the method of rotation at subspace
\cite{QIC, QVQ1, QDCT, QVQ2}.

\textbf{DS5.} Design oracle $O_{d}$ to compute the similarity measurement
value between $c_{i_{0}}$ and $c_{i}$, i.e.,

\begin{equation*}
|\alpha \rangle |c_{i_{0}}\rangle |i\rangle |x_{i}\rangle |c_{i_{k}}\rangle
|0\rangle \overset{O_{c}}{\rightarrow }|\alpha \rangle |c_{i_{0}}\rangle
|i\rangle |x_{i}\rangle |c_{i_{k}}\rangle |d(c_{i},c_{i_{0}})\rangle
\end{equation*}

\textbf{DS6.} Design oracle $O_{f}$ to mark the target patterns:

\begin{equation*}
|\alpha \rangle |c_{i_{0}}\rangle |i\rangle |x_{i}\rangle |c_{i_{k}}\rangle
|d(c_{i},c_{i_{0}})\rangle \overset{O_{f}}{\rightarrow }(-1)^{f(i)}|\alpha
\rangle |c_{i_{0}}\rangle |i\rangle |x_{i}\rangle |c_{i_{k}}\rangle
|d(c_{i},c_{i_{0}})\rangle
\end{equation*}

\ \ \ \ , where $f(i)=\left\{
\begin{tabular}{ccc}
$1$ & $if$ & $0\leq d(c_{i},c_{i_{0}})\leq \alpha $ \\
$0$ & \multicolumn{2}{c}{$otherwise$}%
\end{tabular}%
\right. $.

\textbf{DS6.} Construct pattern recognition\ iteration $G_{pr}$:

\ \ \ \ According to Eq.\ref{eqGeneralGroverIteration},\ $G_{pr}$\ is

\begin{equation*}
G_{pr}=(2|\xi \rangle \langle \xi |-I)(O_{d}O_{c}U_{L})^{\dagger
}O_{f}O_{d}O_{c}U_{L}
\end{equation*}

\textbf{Second, design the following quantum multi-pattern recognition
algorithm to find a target pattern for\ that }$0\leq d(c_{i},c_{i_{0}})\leq
\alpha $\textbf{.}

\textbf{Multi-Quantum Pattern Recognition Algorithm:}

\textbf{Step1.} Initialize $m=1$ and set $\lambda =6/5$. (Any value of $%
\lambda $ strictly between 1 and 4/3 would do.)

\textbf{Step2.} Choose $j$ uniformly at random among the nonnegative
integers smaller than $m$.

\textbf{Step3.} Apply $j$ iterations of $G_{pr}$\ acting on state $|\psi
_{1}\rangle =\frac{1}{\sqrt{N}}|\alpha \rangle |c_{i_{0}}\rangle (\overset{%
N-1}{\underset{i=0}{\sum |i\rangle }})|0\rangle |0\rangle |0\rangle $.

\textbf{Step4.} Observe the 3rd register: let $i_{0}$ be the outcome.

\textbf{Step5.} Calculate value $d(c_{i},c_{i_{0}})$ using classical
computation. If $0\leq d(c_{i},c_{i_{0}})\leq \alpha $, preserve $i_{0}$ and
the signal $x_{i_{0}}$ captured by phased array radar is target pattern, and
exit.

\textbf{Step6.} Otherwise, set $m$ to $min(\lambda m,\sqrt{N})$ and go back
to step 2.

The above quantum multi-pattern recognition algorithm is similar to BBHT
algorithm \cite{BBHT}, which is the improved algorithm of Grover's
algorithm. And the main difference between BBHT algorithm and the presented
algorithm in this paper is that Grover iteration is replaced by pattern
recognition\ iteration $G_{pr}$ that realizes the coupling between quantum
search and the computation of pattern recognition. The above algorithm has
time complexity $O(\sqrt{\frac{N}{M}})$ \cite{BBHT}, where $M$ denotes the
number of target pattern that satisfy the condition $0\leq
d(c_{i},c_{i_{0}})\leq \alpha $. The above presented algorithm is also
suitable to the case in which $N$ is not a big integer according to Long's
research \cite{GLLong}.

\section{Conclusion}

It's the key research topic of signal processing that recognizing
genuine targets real time from the disturbed signal which has giant
amount of data. A quantum algorithm for pattern recognition of
classical signal which has time complexity $O(\sqrt{N})$ is
presented in this paper. Quantum discrete Fourier transform
\cite{QDFT} shows quantum signal processing is possible. This paper
shows that quantum pattern recognition (or quantum image
recognition) for classical signal is also possible.

\begin{acknowledgments}
The first author thanks his teacher prof. G.-C. Guo and the Key Lab. of
Quantum Information, USTC because the first author is brought up from the
lab. The first author thanks prof. Z. F. Han for that he encourages and
helps the first author up till now. The first author thanks assistant prof.
Xudong Huang who is at Harvard Uni. for his interest at the author's
research topic of quantum image compression and quantum image recognition.
The discussion between prof. Huang and the first author makes the first
author decide to open this paper soon.
\end{acknowledgments}


\begin{thebibliography}{99}
\bibitem{Shor} P.W. Shor. \newblock Algorithms for quantum computation
discretelog and factoring. \newblock In \emph{Proc. of the 35th Annual
Symposium on the Foundations of Computer Science}, pages 20--24, Los
Alamitos, CA, 1994. IEEE Computer Society Press.

\bibitem{Grover} Lov~K. Grover. \newblock A fast quantum mechanical
algorithm for database search. \newblock In \emph{Proc. of 28th Annual ACM
Symposium on the Theory of Computing}, pages 212--218, Philadelphia,
Pennsylvania, 1996. ACM Press.

\bibitem{QLS} Chao-Yang Pang. \newblock Loading N-dimensional vector into
quantum registers from classical memory with O(logN) steps. \newblock %
arXiv:quant-ph/0612061, 2006.

\bibitem{QIC} Chao-Yang Pang. \newblock Quantum image compression. \newblock %
Postdoctoral report, Key Laboratory of Quantum Information, University of
Science and Technology of China (CAS), Hefei, China, Jun 2006.

\bibitem{QVQ1} Pang Chao-Yang, Zhou Zheng-Wei, Chen Ping-Xing, and Guo
Guang-Can. \newblock Design of quantum vq iteration and quantum vq encoding
algorithm taking sqrt(n) steps for data compression.
\newblock {\em CHIN.
PHYS.}, 15(3):618--623, 2006.

\bibitem{QDCT} Chao-Yang Pang, Zheng-Wei Zhou, and Guang-Can Guo. \newblock %
Quantum discrete cosine transformation for image compression. \newblock %
arXiv:quant-ph/0601043, 2006.

\bibitem{QVQ2} Pang Chao-Yang, Zhou Zheng-Wei, and Guo Guang-Can. \newblock %
A hybrid quantum encoding algorithm of vector quantization for image
compression. \newblock {\em CHIN. PHYS.}, 15(12):3039--3043, 2006.

\bibitem{QDFT} Chao-Yang Pang and Ben-Qiong Hu. \newblock Quantum Discrete
Fourier Transform with Classical Output for Signal Processing. \newblock %
arXiv:0706.2451, 2007.

\bibitem{HGeometry} Chong-Shan Luo, Chao-Yang Pang, and Yu-Ping Tian. %
\newblock {\em Higher Geometry}. \newblock Higher Education Press, Beijin,
China, 2006.

\bibitem{PR} Shao-Qi Bian, Xue-Gong Zhang, and et.al..
\newblock {\em
Pattern Recognition}. \newblock Tsinghua University Press, Beijin, China,
2000.

\bibitem{BBHT} M.~Boyer, G.~Brassard, P.~Hoyer, and A.~Tap. \newblock Tight
bounds on quantum searching. \newblock arXiv:quant-ph/9605034, 1996.

\bibitem{GLLong} G.~L. Long. \newblock Grover algorithm with zero
theoretical failure rate. \newblock {\em PHYSICAL REVIEW A},
64(2):022307/1--4, 2001.
\end{thebibliography}
\end{document}